\documentclass[preprint,proceedings]{rmaa}

\usepackage{paralist}

\usepackage{psfrag,color}

\def\ie{{i.e.\thinspace}}

\SetYear{2002}
\SetConfTitle{Galaxy Evolution: Theory and Observations}

\title{Forming Globular Cluster Systems Semi-analytically} 

\author{
  M. A. Beasley,\altaffilmark{1} 
  C. M. Baugh,\altaffilmark{2}
  D. A. Forbes,\altaffilmark{1}
  R. M. Sharples,\altaffilmark{2} 
  and C. S. Frenk,\altaffilmark{2}}

\altaffiltext{1}{Astrophysics \& Supercomputing, Swinburne, Australia.}
\altaffiltext{2}{Department of Physics, University of Durham, UK.}

\shortauthor{M. Beasley et al.}
\shorttitle{Globular Clusters}

\fulladdresses{
\item Michael Beasley, Duncan A. Forbes: Astrophysics \&
Supercomputing, Swinburne University, Hawthorn, VIC 3122,
Australia (mbeasley@astro.swin.edu.au).
\item Carlton M. Baugh, Carlos S. Frenk, Ray M. Sharples:
Department of Physics, University of Durham, Durham DH1 3LE, UK.}

\listofauthors{M.A. Beasley, C.M. Baugh, D.A. Forbes,
R.M. Sharples \& C.S. Frenk}
\indexauthor{Beasley, M. A.}
\indexauthor{Baugh, C. M.}
\indexauthor{Forbes, D. A.}
\indexauthor{Sharples, R. M.}
\indexauthor{Frenk, C. S.}

\abstract{We describe a scheme for the formation of globular
cluster systems in early-type galaxies using a 
semi-analytic model of galaxy formation. 
Operating within a $\Lambda$CDM cosmology,
we assume that metal-poor globular clusters are formed at high-redshift
in pre-galactic fragments, and that the subsequent gas-rich merging
of these fragments leads to the formation of metal-rich 
clusters. We compare our results with contemporary data,
and look at the particular case of the
globular cluster and stellar metallicity distribution function
of the nearby elliptical Centaurus A.}

\resumen{Describimos el escenario para la formacion de sistemas de cumulos
globulares en galaxias elipticas utilizando un modelo semi-analitico
de formacion de galaxias.
Trabajando en una cosmologia de tipo $\Lambda$CDM, asumimos que los
cumulos globulares de baja metalicidad se forman a altos redshifts en
fragmentos pre-galacticos, y que la consiguiente fusion rica en gas de
estos fragmentos conduce a la formacion de los cumulos globulares de alta
metalicidad. Comparamos nuestros resultados con datos
contemporaneos y estudiamos el caso particular de la funciones de
distribucion de metalicidad estelar y de los cumulos globulares
de la galaxia eliptica cercana Centaurus A.}

\addkeyword{Stars: Star Clusters}
\addkeyword{Galaxies: Elliptical}

\begin{document}
\maketitle

\section{General}
\label{sec:intro}

Globular clusters (GCs) are long-lived tracers  
star formation, ranging from ancient, metal-poor
stellar populations (e.g., Galactic halo GCs) to young, 
solar-metallicity stellar populations created during
interactions (e.g., Schweizer \etal\@ 1996). Due to their relatively 
``simple'' nature (they are idealised simple stellar 
populations), the ages and metallicities of GCs can in
principle be unambiguously determined. Hence, the study of GC systems
affords a unique insight into the formation histories of galaxies.

We have investigated the formation of the GC systems 
of elliptical galaxies using the fiducial semi-analytic model of
Cole \etal\@ (2000). 
We (Beasley \etal\@ 2002a)
assume GC formation occurs in two modes;
in pre-galactic fragments with GC formation 
truncated at high redshift, and during the dissipative merging of 
these fragments. 

\begin{figure}
  \includegraphics[width=\columnwidth]{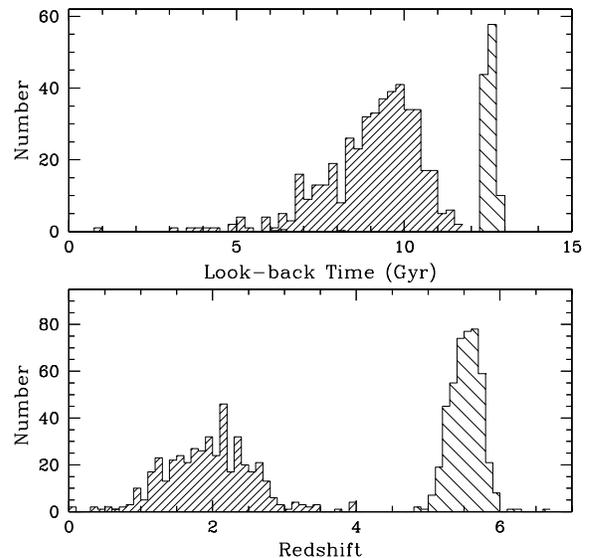}
  \caption{The distribution in mean ages
for 450 model GC systems. 
{\it Top panel:} Distribution in look-back
time for the metal-rich GCs (dark histogram)
and metal-poor GCs (light histogram). 
The metal-poor GC histogram has been scaled down
by a factor of 6 for display purposes.
{\it Bottom panel:} Distribution in formation 
redshifts for the GC sub-populations.}
  \label{fig:one}
\end{figure}

With these assumptions, we produced 450 realisations 
of elliptical galaxy GC systems of over a range of halo masses 
\ie, 1.0$\times 10^{13} h^{-1} M_\odot \leq$ M$_{\rm h} 
\leq 1.3\times 10^{15} h^{-1} M_\odot$.


\begin{figure}
  \includegraphics[width=\columnwidth]{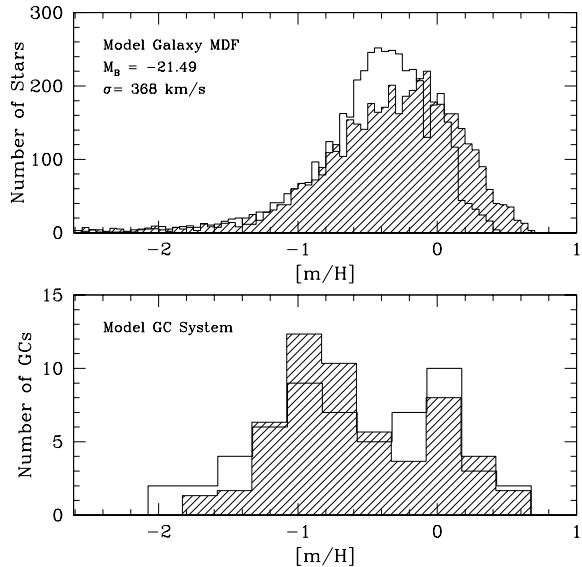}
  \caption{{\sl Upper panel:}  Comparison between the semi-analytic model
stellar metallicity distribution function 
(solid line), and the observed MDF
of old stars in NGC 5128 from Harris \& Harris (2002; shaded histogram).  
{\sl Lower panel:}  Comparison between the model 
GC MDF (solid line) and the 
observed NGC~5128 GCs (Held \etal\@ 2002 private comm; shaded histogram).}
  \label{fig:two}
\end{figure}

Using simple stellar population models 
(Kurth, Fritz-v. Alvensleben \& Fricke 1999), we can transfer our
predictions into the observable plane. 
We have compared our model results to contemporary data 
for the GC systems of elliptical galaxies, 
and the principle results 
from our study are as follows:


{\it (i)} the total number of GCs (N) scales with host galaxy luminosity
as N $\propto L^{1.25}$, rather than the expected N $\propto L^{1.0}$ for 
constant GC formation efficiencies. This scaling is very similar
to the observational data, and in the model is due to a 
$\mathcal{M}/L \propto L$ dependence for $L_*$ ellipticals.

{\it (ii)} the formation of metal-rich GCs in low-redshift
mergers does not significantly alter the S$_{\rm N}$ (number of GCs per
unit starlight) of the host elliptical. This reflects the 
small gas-fractions of merger progenitors at later epochs. 

{\it (iii)} the metal-poor GCs have {\it mean} ages of
$\sim$ 12 Gyr, whilst the metal-rich GCs have {\it mean}
ages of $\sim$ 9 Gyr (e.g. Fig~1). 
However, the age range of the metal-rich GCs 
in many individual galaxies
is significant (5 $\sim 12$ Gyr) with a small fraction 
forming at the present epoch. 

\adjustfinalcols

Our results suggest that metal-rich GCs may directly trace the 
merger history of their host galaxy whenever
star formation occurs. Therefore, the metal-rich GCs are 
remnants of both star formation at later times, and 
the dynamical assembly of their host galaxies, possibly
providing a key to probing the formation epoch of 
elliptical galaxies.

Furthermore, our model provides explicit
information on the ages and abundances of both the GCs 
and the ``field'' stars which 
aggregate to produce the final galaxy.
Recently,  Harris \& Harris (2002) have constructed
a metallicity distribution function (MDF) for the
nearest giant elliptical galaxy Centaurus A (NGC~5128) 
using {\it HST}.
By selecting model ellipticals occupying similar environments
to this galaxy, we can
directly compare its predicted and observed stellar
content (see Beasley \etal\@ 2002b).

Such a comparison is shown in Fig~2 for the stellar
MDF of NGC~5128 and its GC system.
The model MDFs are qualitatively similar to those observed; 
both model and data have stellar components which are 
predominantly metal-rich ($\sim$ 0.8$Z_{\odot}$), but
possess a small number of metal-poor stars extending
down to 0.002$Z_{\odot}$. Early gas infall
in the model ameliorates the so-called ``G-dwarf'' 
problem (e.g. Kauffmann 1996).

Interestingly, we find that the model MDFs harbour 
a greater fraction of stars at $Z > Z_{\odot}$ than the observations, 
yielding a broader (by $\sim$ 0.1 dex), more metal-rich MDF.
This is possibly a result of the fact that these outer-bulge 
observations are missing some of the
highest-metallicity stars in NGC~5128.
Such a comparison facilitates the possible reconstruction of
the star formation history of this nearby elliptical galaxy.

\end{document}